# Evolution of an Early Titan Atmosphere: Comment

R.E. Johnson[1,2], O.J. Tucker[3], A.N. Volkov[4]
[1] Engineering Physics, University of Virginia, Charlottesville, VA 22904
[2] Physics Department, New York University, NY, NY
[3] Department of Atmospheric Oceanic and Space Sciences, University of Michigan
[4] Department of Mechanical Engineering, University of Alabama, Tuscaloosa AL 35487

**Abstract**
Escape of an early atmosphere from Titan, during which time $NH_3$ could be converted by photolysis into the present $N_2$ dominated atmosphere, is an important problem in planetary science. Recently Gilliam and Lerman (2014) estimated escape driven by the surface temperature and pressure, which we show gave loss rates that are orders of magnitude too large. Their model, related to Jeans escape from an isothermal atmosphere, was used to show that escape driven only by surface heating would deplete the atmospheric inventory of N for a suggested Titan accretion temperature of ~355 K. Therefore, they concluded that the accretion temperature must be lower in order to retain the present Titan atmosphere. Here we show that the near surface atmospheric temperature is essentially irrelevant for determining the atmospheric loss rate from Titan and that escape is predominantly driven by solar heating of the upper atmosphere. We also give a rough estimate of the escape rate in the early solar system ($\sim 1.0 \times 10^4$ kg/s) consistent with an inventory of nitrogen being available over the time period suggested by Atreya et al. (1978) for conversion of $NH_3$ into $N_2$.

**Introduction**

The evolution of and escape from Titan's atmosphere is still debated in spite of the large amount of Cassini data on Titan's upper atmosphere: ~100 passes through the atmosphere. There is agreement that the present escape rate for $H_2$ (e.g., Tucker et al. 2013) is roughly consistent with the rate of photolysis of $CH_4$ and the subsequent precipitation of larger carbon containing molecules to Titan's surface (e.g., Atreya et al. 2006). However, the present escape rate for carbon and nitrogen containing molecules is small and still debated (e.g., Tucker et al. 2013). Although the surface temperature (~94K) is much lower than estimates of the temperature following accretion (>~300K) used in Gilliam and Lerman (2014) (hereafter GL), it is known from modeling of Titan (Johnson et al. 2009a,b) and other planetary bodies with relatively thick atmospheres, that the escape rate is *not* very sensitive to the near surface atmospheric temperature.

GL used the near surface temperature of the atmosphere and assumed thermal escape occurred from approximately one scale height above the surface, rather than from the exobase radius, which at present is $r_x \sim 1.6\ r_0$ and was much higher early in its evolution. A thick atmosphere *only* heated at the surface decreases in temperature and pressure with altitude, so that the thermal escape rate from the exobase region can be orders of magnitude smaller than that estimated using the near surface temperature (Volkov et al. 2011a,b). Their approach, as outlined in their Tables 7-9, is related to the well-known Jeans expression for the globally-averaged molecular loss rate, $\Phi_J$, for thermally driven escape:

$$\Phi_J = \pi r^2 n(r) <v> [1+\lambda(r)] \exp[-\lambda(r)] \qquad \text{Eq. 1}$$

where $n(r)$, $<v>$, and $\lambda(r)$ are molecular density, mean thermal speed and Jeans parameter in the atmosphere at a radius $r$ from the center of the planet. The Jeans parameter is $\lambda(r) = U(r)/kT(r)$, the ratio of the gravitational binding energy of a molecule, $U(r)$, to a measure of its thermal energy, $kT$, with $k$ the Boltzmann constant and $T(r)$ the local temperature. Eq.1 is typically evaluated at the exobase radius ($r = r_x$). For a *thin* atmosphere, an estimate is often made replacing $r_x$ by the surface radius, $r_0$, referred to as the surface-Jeans approximation (e.g., Volkov et al. 2011a, b). For the early Titan atmosphere in GL at 355 K, with 5.8 bar of $NH_3$ and a surface Jeans parameter $\lambda_0 = 20.1$, and 19.6 bar of $CH_4$ with $\lambda_0 = 18.9$, using $r_0 = 2576$km in Eq. 1 gives an upward flux of molecules with sufficient energy to escape ~ $0.65 \times 10^{35}(NH_3)$/s and ~ $6.8 \times 10^{35}(CH_4)$/s. The rough approximations in GL at one scale height above the surface is about a factor of two larger: ~$1.4 \times 10^{35}(NH_3)$/s and $15. \times 10^{35}(CH_4)$/s respectively. For the atmosphere of interest these are *both* very poor estimates of the actual escape rates as described below.

**Escape Driven by Surface Heating**

Molecular kinetic simulations of escape have shown that the escape rate can be simulated reasonably well for a *thick* atmosphere by applying Eq.1 iteratively as an upper boundary condition in a fluid dynamics simulation (Erwin et al. 2013), and that these results can be roughly scaled using planetary parameters (Volkov et al 2011b). Such simulations, often referred to as fluid-Jeans simulations, supplemented by molecular kinetic simulations, were carried out for nitrogen atmospheres ignoring direct atmosphere heating and spanning a range of surface values of the Jeans parameter ($\lambda_0 = 10$ to 30) and surface column densities ($N_0 = 10^{14}$ to $10^{26}$ $N_2$/cm$^2$) as shown in Fig. 2 of Johnson et al.

(2015). Here $N_0 = P_{vap}/mg_0$, with $P_{vap}$ the vapor pressure and $g_0 = U(r_0)/r_0$ the surface gravity. The ratio, $R$, of the simulated escape rate divided by the SJ rate in Eq. 1 were roughly fit as $R^{-1} = R_1^{-1} + R_2^{-1}$, where $R_1 \sim 1/Kn_0^{0.09}$ applied at the smaller $N_0$ and $R_2 \sim 70 [Kn_0 \exp(\lambda_0)]/\lambda_0^{2.55}$ applies at the larger $N_0$. Here $Kn_0$ is the surface value of the Knudsen number used here as $Kn_0 = (\lambda_0 N_0 \sigma_{eff})^{-1}$ with $\sigma_{eff}$ the effective collision cross section between the atmospheric molecules. Using $Kn_0$ for an early Titan atmosphere with an average molecular mass, $\sim 16.2$amu, for the relevant $\lambda_0$ and surface pressure, $r_x$ is many times Titan's radius and the thermally driven mass loss rate is $\sim 10^{-8}$ *smaller than the estimate in GL*. This large difference indicates that neither the Jeans escape rate applied at the surface nor the rough approach developed by GL can be used to calculate escape rate from an early Titan atmosphere. If the near surface atmospheric temperature determined the escape rate, then, contrary to the conclusion in GL, the temperature would have to be *much* higher than 355 K not smaller.

**Ecape Driven by Heating of the Upper Atmosphere**

Although escape rates from an early Titan atmosphere are large, they are not very sensitive to near surface atmospheric temperature. Rather, the energy deposited in the upper atmosphere dominates the loss rate for thick atmospheres. Escape due to direct heating of the upper atmosphere was recently modeled using molecular kinetic simulations for Titan and Pluto (Tucker and Johnson 2009; Tucker et al. 2012; 2013; Erwin et al. 2013). The relevant energy can be deposited by the ambient plasma (e.g., Johnson 2004) or by the short wavelength solar flux (Lammer et al., 2008). Molecular kinetic simulations for the UV-EUV driven escape from Titan's *present* 1.5 bar atmosphere have lead to very small rates for loss of heavy particles ($N_2$, $CH_4$: Tucker and Johnson 2009; Tucker et al. 2013) that cannot be scaled. Of course, an accurate description of the loss rate by absorption of the UV solar flux requires detailed knowledge of the molecular physics. However, using molecular kinetic simulations we have recently confirmed that a rough approximation, called the energy limited escape rate, can be used over a board range of solar fluxes and object sizes (Erwin et al. 2013; Johnson et al. 2013). In this approximation the globally averaged mass loss rate, *(dM/dt)*, is the energy limited rate, $(dM/dt)_{EL}$:

$$(dM/dt) \sim c\, (dM/dt)_{EL} \sim c\, [m\, Q\, /\, U(r_a)] \qquad (2)$$

Here $m$ is the molecular mass, $Q$ is the globally-averaged heating rate of the *upper* atmosphere, $U(r)$ is the gravitational binding energy of a molecule at a depth, $r_a$, assumed to be below the energy absorption peak and $c$ accounts for other effects but is often assumed to be approximately one. The expression in Eq. 2 with $c \sim 1$ is reasonable when adiabatic cooling by molecular outflow dominates in the upper atmosphere but breaks down, not surprisingly, at small heating rates and, less obviously, when the gas goes sonic below the exobase (Johnson et al. 2013). The latter is the case because the energy per molecule carried off increases rapidly with increased heating and the flux is limited by the supply from below $r_a$. More detailed expressions for $(dM/dt)_{EL}$ have been used which include estimates of IR absorption and cooling (e.g., Erwin et al. 2013).

The present globally averaged heat flux deposited in Titan's atmosphere is primarily due to UV absorption and photolysis of $CH_4$: $\sim 1.4 \times 10^{-5}$J/m$^2$/s for average solar conditions using an estimated heating efficiency (Krasnopolsky 2009). In the early solar system when the sun is more active, the relevant UV flux is enhanced by $\sim (4.56 Gyr/t)^{0.72}$ where $t$ is the time from formation (Ribas et al. 2005). This increases the deposited heat flux at ~ 0.1Gyr by a factor of ~15.6 [e.g., ~ $2.2\times10^{-4}$ J/m$^2$/s]. Solar radiation also dissociates NH$_3$ with a cross-section greater than $5\times10^{-23}$ cm$^2$ below 2300A where the globally average solar flux incident on Titan is ~ $3.5\times10^{11}$ cm$^{-2}$ s$^{-1}$ (Atreya et al. 1978). Therefore, longer solar wavelengths can contribute significantly to the heating. But this occurs much deeper into Titan's atmosphere, as can be seen by comparing the cross section above to the CH$_4$ absorption cross section for Lyman-α absorption, ~$1.8\times10^{-17}$ cm$^2$. Since the early solar wind was also more intense so that Titan likely spent much more time orbiting outside Saturn's magnetopause, solar-wind plasma-induced heating could also be relevant. Here, we use the heat flux above as a rough lower bound to estimate the globally averaged heating rate: $Q = [2.2\times10^4 J/m^2/s](4\pi r_a^2)$, where $r_a$ is an estimate of the effective radius of the disc above which the relevant UV radiation is fully absorbed.

Scaling the simulations reported in Johnson et al. (2013; 2015) to the relevant $\lambda_0$ and pressure, the heated upper atmosphere becomes highly extended, e.g. $r_x$ ~16 $r_T$ where $r_T$ is the radius of Titan ($r_0$ ~2576km), with the Lyman-α fully absorbed above ~ $10^{18}$ CH$_4$/cm$^2$ ($r > r_a$ ~ 5). Using the enhancement in the UV flux at 0.1Gyr and assuming it is fully absorbed above $r_a$ ~ $5r_0$, $Q$ ~ [$3.3\times10^{11}$ J/ s] and $U(r_a)/m$ ~ $7.\times10^5$ J/kg. Therefore, based on Eq. 2, $(dM/dt)_{EL}$ ~ $5.0\times10^5$ kg/s. The ratio of this heating rate to the lower bound for the criterion for sonic escape, $Q_c$ derived in Johnson et al. (2013), gives $Q/Q_c$ ~18. For such values the escape rate goes sonic well below the exobase. As seen from Fig. 2 in Johnson et al (2013), the heat carried off per molecule increases nearly linearly for large $Q$ as the flux of escaping molecules is limited by the supply from below. Based on this ratio, the rate in Eq. 2 drops by a factor $c$ ~ 0.02 so that $(dM/dt)$ ~ $1.0\times10^4$ kg/s is a rough bound on the loss rate from the proposed early Titan atmosphere. Therefore, although this estimate of the upper atmospheric heating rate is very rough, when $Q \gg Q_c$ the molecular loss rate is somewhat insensitive to the size of $Q$ so that the molecular loss rate would be roughly the same, $(dM/dt)$ ~ $10^4$ kg/s, even ignoring the early solar system flux enhancement. Although these loss rates are robust, they are orders of magnitude smaller than the $1.5\times10^{10}$ kg/s at 300 K which GL required to reduce the N content to the present value in a time much less than 0.1Gyr.

Atreya et al. (1978) estimated that NH$_3$ can be converted to N$_2$ in ~ 0.16Gyr, if the atmospheric temperature remains above ~150K during that time period. Using ~ $10^4$ kg/s and assuming the atmosphere remains well mixed and warm, then in ~ 0.16 Gyr the N content would decrease by 2-3 bars due to escape of NH$_3$ and the loss of the H associated with the conversion of the remaining NH$_3$ to N$_2$. Even considering the approximate nature of the above estimates, it is clear that the source of present 1.5 bar N$_2$ atmosphere at Titan could easily be an early NH$_3$ atmosphere as has been discussed (e.g., Lammer et al. 2008) and is not strongly affected by differences in the surface temperature. Therefore, much more detailed simulations of Titan's atmospheric escape in the early solar system are needed and are in progress.

**Acknowledgements**
REJ acknowledges support from NASA's ROSES PATM program, from the NASA's CASSINI mission through SwRI; REJ & ANV acknowledge support from NASA's ROSES OPR program; OJT acknowledges support from the grant NNH12ZDA001N from the NASA's ROSES OPR program.

**References**

Atreya, S.K., Donahue, T.M., Kuhn, W.R., 1978. Evolution of a nitrogen atmosphere on Titan. Science 201, 611–613.

Atreya, S.K., Adams, E.Y., Niemann, H.B., Demick-Montelara, J.E., Owen, T.C., Fulchignoni, M., Ferri, F., Wilson, E.H., 2006. Titan's methane cycle. Planet. Space Sci. 54, 1177–1187.

Erwin, J., Tucker, O.J., and R.E. Johnson, R.E., 2013. Hybrid fluid/kinetic modeling of Pluto's atmosphere, Icarus 226, 375-384 (2013).

Johnson, R.E., 2004. The Magnetospheric Plasma-Driven Evolution of Satellite Atmospheres, Astrophys. J. 609: L99-L102.

Johnson, R.E., Volkov, A.N. and J.T. Erwin, J.T., 2013. Molecular-kinetic Simulations of Escape from the Ex-planet and Exoplanets: Criterion for Transonic Flow, ApJ. Lett. 768, L4 (6pp); Erratum: ApJL 774, 90 (1pp).

Johnson, R.E., A. Oza, L. A. Young, A. N. Volkov and C. Schmidt 2015. Volatile Loss and Classification of Kuiper Belt Objects. Astrophys. J. in press (**arXiv:1503.05315**)

Lammer, H., J.F. Kasting, E. Chassefière, R.E. Johnson, Y.N. Kulikov, F.Tian, 2008.Atmospheric Escape and Evolution of Terrestrial Planets and Satellites Space Sci Rev. 139, 399–436

Krasnoplosky, V.A., 2009 A photochemical model of Titan's atmosphere and ionosphere Icarus, 201, 226-256

Tucker, O.J., and R.E. Johnson 2009. Thermally driven atmospheric escape: Monte Carlo simulations for Titan's atmosphere, Planetary Space Sci. 57, 1889-1894.

Tucker, O.J., J.T. Erwin, J.I. Deighan, A.N. Volkov, and R.E. Johnson, 2012. Thermally driven escape from Pluto's atmosphere: A combined fluid/kinetic model, Icarus 217, 408-415.

Tucker, O.J., R.E. Johnson, J.I. Deighan, A.N. Volkov 2013. Diffusion and thermal escape of $H_2$ from Titan's atmosphere: Monte Carlo Simulations, Icarus 222, 149-158.

Volkov, A.N., R.E. Johnson, O.J. Tucker, and J.T. Erwin, 2011a. Thermally driven atmospheric escape: Transition from hydrodynamic to jeans escape, The Astrophys. J. Lett. 729: L24 (5pp).

Volkov, A.N, O.J. Tucker, J.T. Erwin, and R.E. Johnson, 2011b. Kinetic simulations of thermal escape from a single component atmosphere, Phys. of Fluids 23, 066601.